\documentclass[usenatbib]{mn2e}
\usepackage[dvips]{graphicx}
\usepackage{amssymb}
\usepackage{txfonts}
\usepackage{colordvi}

\newcommand{\integral}{{\textit{INTEGRAL}}}
\newcommand{\xte}{{\textit{RXTE}}}

\newcommand{\sax}{{\textit{Beppo\-SAX}}}
\newcommand{\gro}{{\textit{CGRO}}}
\newcommand{\fermi}{{\textit{Fermi}}}
\newcommand{\agile}{{\textit{AGILE}}}
\newcommand{\swift}{{\textit{Swift}}}
\newcommand{\msun}{{\rm M}_{\sun}}
\newcommand{\rsun}{{\rm R}_{\sun}}

\newcommand{\g}{$\gamma$}

\let\oldhat\hat

\renewcommand{\hat}[1]{\oldhat{\mathbfit{#1}}}

\newbox\grsign \setbox\grsign=\hbox{$>$} \newdimen\grdimen \grdimen=\ht\grsign
\newbox\simlessbox \newbox\simgreatbox \newbox\simpropbox
\setbox\simgreatbox=\hbox{\raise.5ex\hbox{$>$}\llap
     {\lower.5ex\hbox{$\sim$}}}\ht1=\grdimen\dp1=0pt
\setbox\simlessbox=\hbox{\raise.5ex\hbox{$<$}\llap
     {\lower.5ex\hbox{$\sim$}}}\ht2=\grdimen\dp2=0pt
\setbox\simpropbox=\hbox{\raise.5ex\hbox{$\propto$}\llap
     {\lower.5ex\hbox{$\sim$}}}\ht2=\grdimen\dp2=0pt
\def\ga{\mathrel{\copy\simgreatbox}}
\def\la{\mathrel{\copy\simlessbox}}

\title[High-energy gamma-rays from Cyg X-1]{High-energy gamma-ray emission from Cyg X-1 measured by \textit{Fermi}\/ and its theoretical implications}

\author[D. Malyshev, A. A. Zdziarski and M. Chernyakova]
{Denys Malyshev,$^1$ Andrzej A. Zdziarski$^2$ and Maria Chernyakova$^{3,4}$ \\
$^1$Bogolyubov Institute for Theoretical Physics, Metrologichna str., 14-b, Kiev 03680, Ukraine\\
$^2$Centrum Astronomiczne im.\ M. Kopernika, Bartycka 18, PL-00-716 Warszawa, Poland\\
$^3$School of Physical Sciences, Dublin City University, Glasnevin, Dublin 9, Ireland \\ 
$^4$DIAS, Fitzwiliam Place 31, Dublin 2, Ireland\\
}

\date{Accepted 2013 June 24.  Received 2013 June 22; in original form 2013 May 25}

\pagerange{\pageref{firstpage}--\pageref{lastpage}}
\pubyear{2013}

\begin{document}

\maketitle

\label{firstpage}

\begin{abstract}
We have obtained measurements and upper limits on the emission of Cyg X-1 in the photon energy range of 0.03--300 GeV based on observations by \fermi. We present the results separately for the hard and soft spectral states, as well for all of the analysed data. In the hard state, we detect a weak steady emission in the 0.1--10 GeV range with a power-law photon index of $\Gamma\simeq 2.6\pm 0.2$ at a $4\sigma$ statistical significance. This measurement, even if considered to be an upper limit, strongly constrains Compton emission of the steady radio jet, present in that state. The number of relativistic electrons in the jet has to be low enough for the spectral components due to Compton upscattering of the stellar blackbody and synchrotron radiation to be within the observed fluxes. If optically-thin synchrotron emission of the jet is to account for the MeV tail, as implied by the recently-claimed strong polarization in that energy range, the magnetic field in the jet has to be much above equipartition. The GeV-range measurements also strongly constrain models of hot accretion flows, most likely present in the hard state, in which \g-rays are produced from decay of neutral pions produced in collisions of energetic ions in an inner part of the flow. In the soft state, the obtained upper limits constrain electron acceleration in a non-thermal corona, most likely present around a blackbody accretion disc. The coronal emission above 30 MeV has to be rather weak, which is most readily explained by absorption of \g-rays in pair-producing photon-photon collisions. Then, the size of the bulk of the corona is less than a few tens of the gravitational radii. 
\end{abstract}
\begin{keywords}
acceleration of particles -- accretion, accretion discs -- gamma-rays: general -- gamma-rays: stars -- stars: individual: Cyg~X-1 -- X-rays: binaries.
\end{keywords}

\section{Introduction}
\label{intro}

Cyg X-1 is an archetypical and widely studied black-hole binary, discovered in 1964 \citep{bowyer65}. It shows two main spectral states, hard and soft (see, e.g., \citealt{zg04} for a review). Most of the time, it is found in the hard state. In that state, the main component of its X-ray spectrum appears to be due to thermal Comptonization in a plasma with the electron temperature of $k T_{\rm e}\sim 100$ keV, which shows a sharp cutoff [in $EF(E)$] at energies $E\ga 200$ keV. In addition, there is a clear high-energy tail on top of that spectrum, measured up to $\sim 3$ MeV (e.g., \citealt{mcconnell02}, hereafter M02; \citealt*{jrm12,zls12}, hereafter ZLS12). The origin of the photon tail may be Compton scattering by a power-law component beyond the thermal electrons in the accretion flow (e.g., M02). On the other hand, the emission in the tail, between $\sim$0.2--0.4 MeV and $\sim$1--3 MeV, has been recently claimed to be strongly polarized \citep{l11,jourdain12}. If this is the case, that emission has to be due to a high-energy tail of optically-thin synchrotron jet emission. The jet will also emit numerous high-energy \g-rays via Compton upscattering of both synchrotron and stellar photons (see, e.g., \citealt*{atoyan99,Georganopoulos02}), which prediction can be confronted with observations. The accretion flow itself may also emit \g-rays via decay of pions produced by collisions of energetic ions. There have been reported upper limits in the hard and soft states at $\geq 100$ MeV from \agile\/ \citep{sabatini10,sabatini13}. However, their refinement is obviously of great interest for constraining the physics of both the jet and the accretion flow.

In the soft state, there is a strong disc blackbody component in the X-ray spectrum, peaking [in $EF(E)$] at $\sim 1$ keV, followed by a pronounced high-energy tail. The best-studied occurrence of the soft state in Cyg X-1 is that of 1996 (\citealt{gierlinski99}; M02). In that case, the high-energy tail had the photon index of $\Gamma\simeq 2.5$ and it extended up to $\sim 10$ MeV. The energy up to which the soft-state tail extends has remained unknown, and measurements and upper limits at energies $> 10$ MeV can constrain the nature of its source, usually thought to be a corona above an inner part of an accretion disc. 

Here, we present upper limits and measurements of steady high-energy \g-ray emission of Cyg X-1 measured by the Large Area Detector (LAT) on board of \fermi. We compare the limits with predictions of theoretical models. In Section \ref{accretion}, we consider high-energy tails predicted by accretion models. In Section \ref{jet}, we consider predictions of jet models. We take into account emission in the GeV band predicted by jet synchrotron self-Compton and Comptonization of blackbody radiation from the donor. 

The orbital period of Cyg X-1 is $P\simeq 5.6$ d. The masses of the components, the radius and temperature of the donor, and the inclination of the binary, $i$, still remain somewhat uncertain. Based on Zi{\'o}{\l}kowski (2005, 2013), \citet{cn09} and \citet{orosz11}, we adopt the black-hole mass of $M_{\rm X}\simeq 16\msun$, the mass of the donor of $M_*\simeq 27\msun$, its radius and effective temperature of $R_*\simeq 19\rsun$, $T_*\simeq 2.8\times 10^4$ K, respectively, and $i\simeq 29\degr$. (Here $\msun$ and $\rsun$ are the solar mass and radius, respectively.) These parameters correspond to the stellar luminosity of $L_*\simeq 8\times 10^{38}$ erg s$^{-1}$ and the separation between the components of $a\simeq 3.2\times 10^{12}$ cm. We adopt the distance to Cyg X-1 of $D=1.86$ kpc \citep{reid11}. The opening angle of the steady jet, present in the hard state, is taken as $\Theta_{\rm j}= 2\degr$ \citep{stirling01}, and its velocity as $\beta_{\rm j}= 0.6$ \citep*{stirling01,gleissner04,mbf09}. 

\label{fermi}
\begin{figure*}
\centerline{\includegraphics[width=8cm]{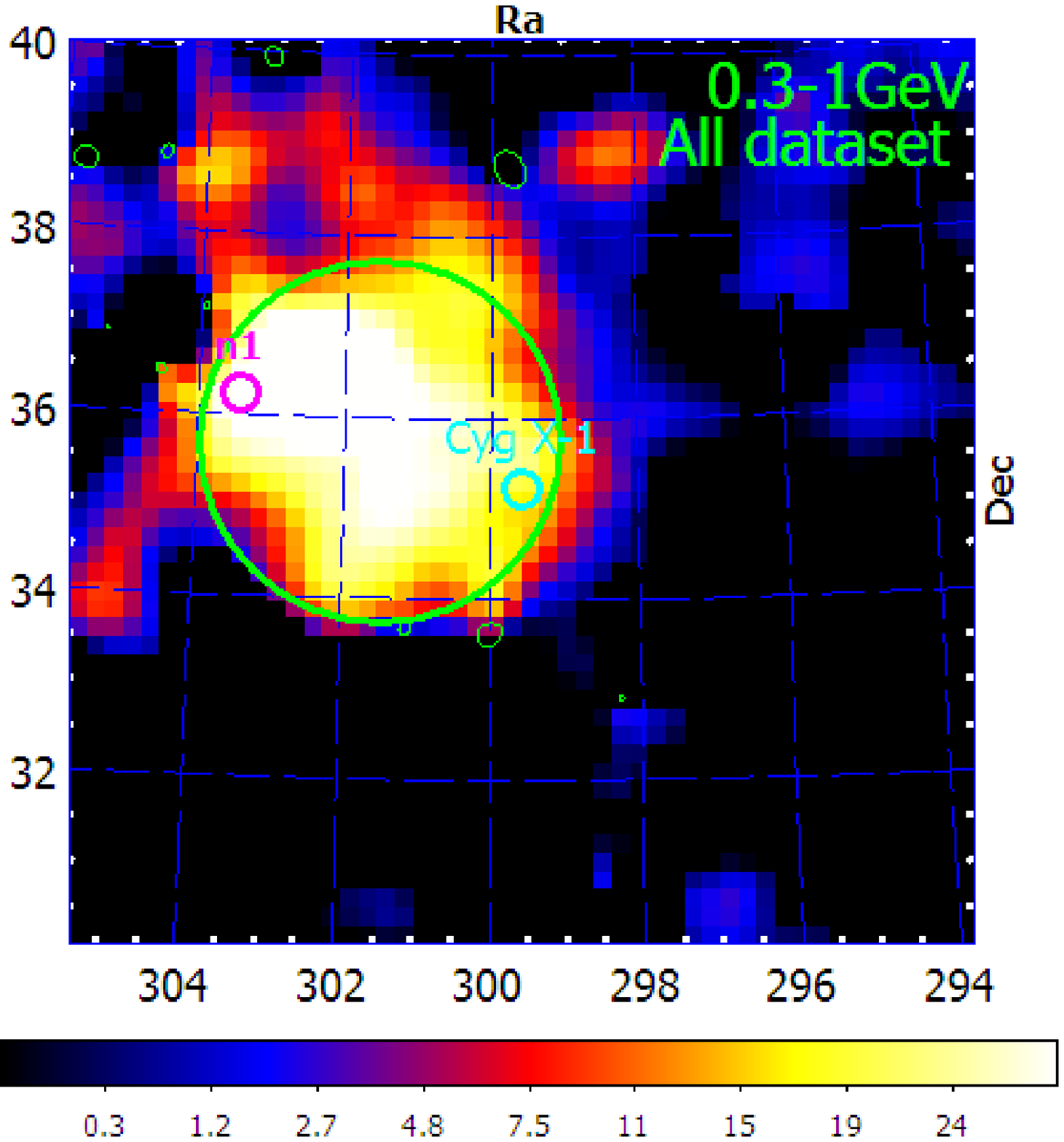} \includegraphics[width=8cm]{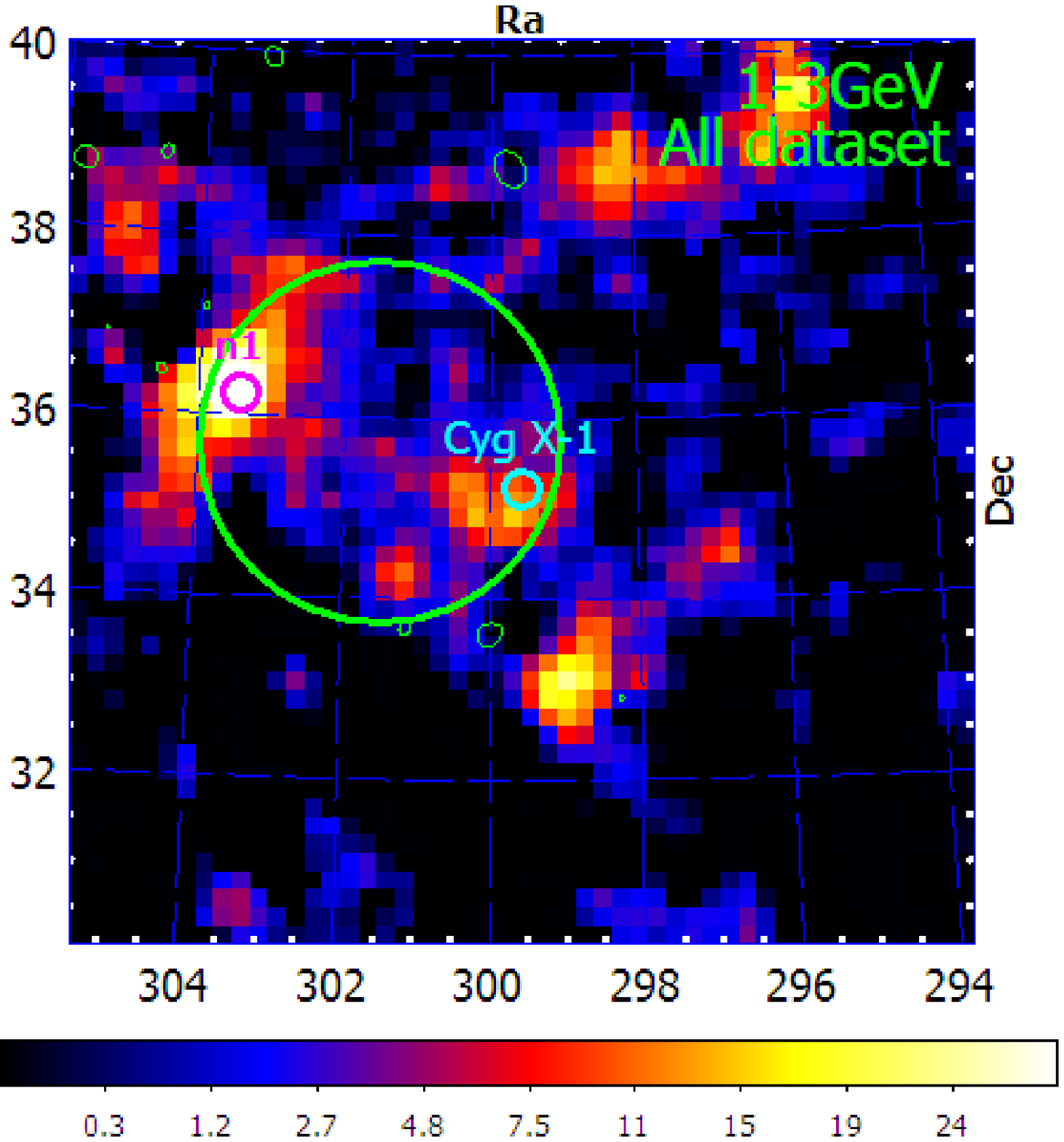}}
\centerline{\includegraphics[width=8cm]{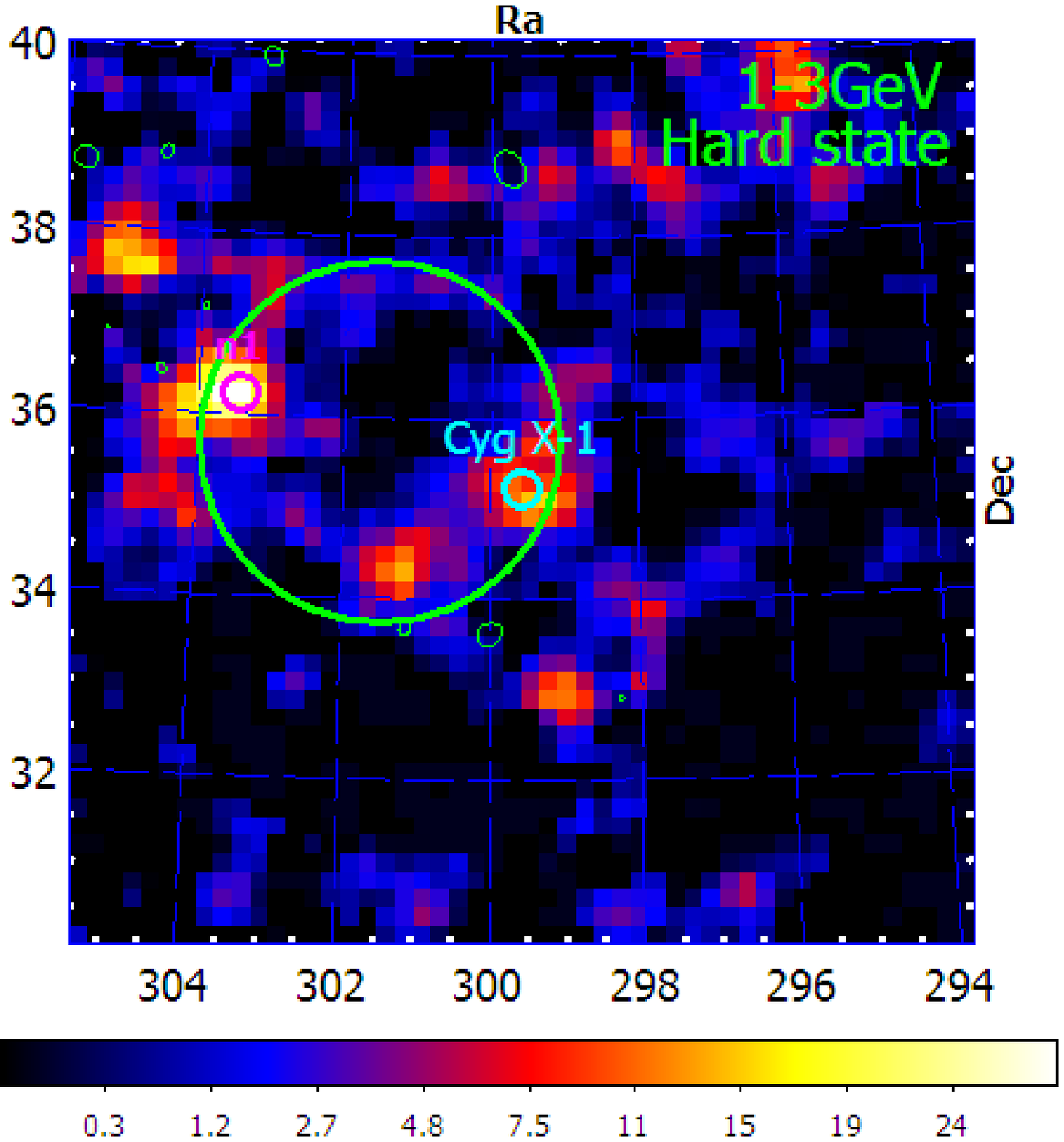} \includegraphics[width=8cm]{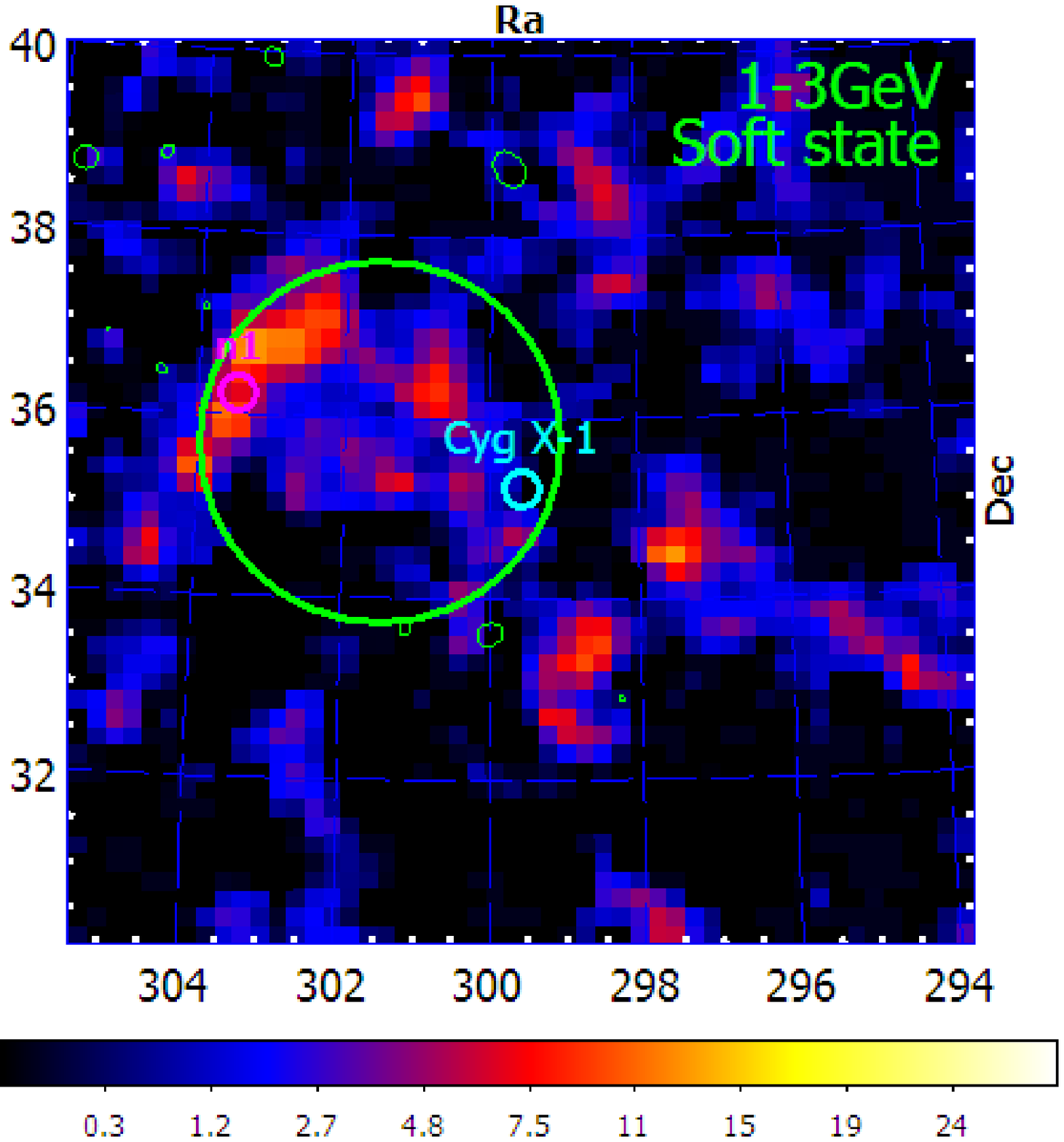}}
\caption{Test-statistic maps of the $10\degr \times 10\degr$ region around the position of Cyg X-1 with the pixel size of $0.2\degr$. Only sources from the 2-yr \fermi\ catalogue (shown with small green ellipses) were subtracted from the maps. The top panels are for the 0.3--1~GeV (left) and 1--3~GeV (right) data. The additional diffuse source (identified with Cyg OB3 association) and the source {\tt n1} (possibly corresponding to SNR G073.9+00.9) are marked with the solid green and magenta circles, respectively. The bottom panels show 1--3~GeV the data split into (left) the hard and (right) soft states. The significance of a point source can be estimated as $\sqrt{TS}$.}
\label{ts_maps}
\end{figure*}

\begin{figure*}
\centerline{\includegraphics[width=13cm]{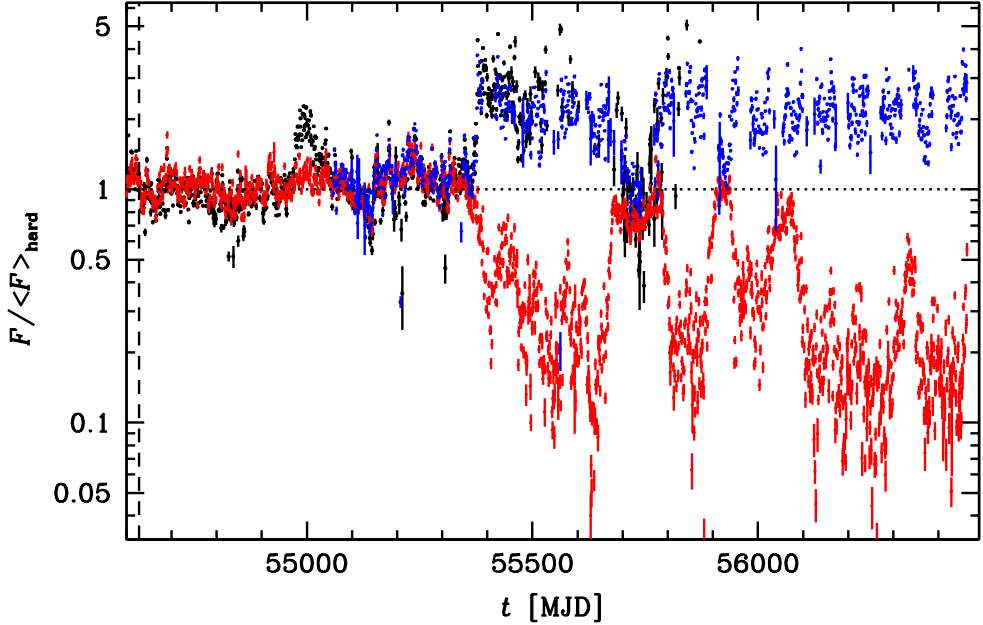}} 
\caption{Light curves of Cyg X-1 from the ASM (1.5--12 keV, black error bars), BAT (15--50 keV, red error bars), and MAXI (2--20 keV, blue error bars) normalized to their respective average hard-state values (dotted line). The adopted ASM, BAT and MAXI average rates are $\langle F\rangle \simeq 20.7$ s$^{-1}$, $\simeq 0.173$ cm$^{-2}$ s$^{-1}$, and $\simeq 1.0$ cm$^{-2}$ s$^{-1}$, respectively. The vertical dashed line denotes the launch of \fermi.
} \label{lc}
\end{figure*}

\section{Analysis of the \textit{FERMI}\/ data}

\begin{figure*}
\centerline{\includegraphics[width=12cm]{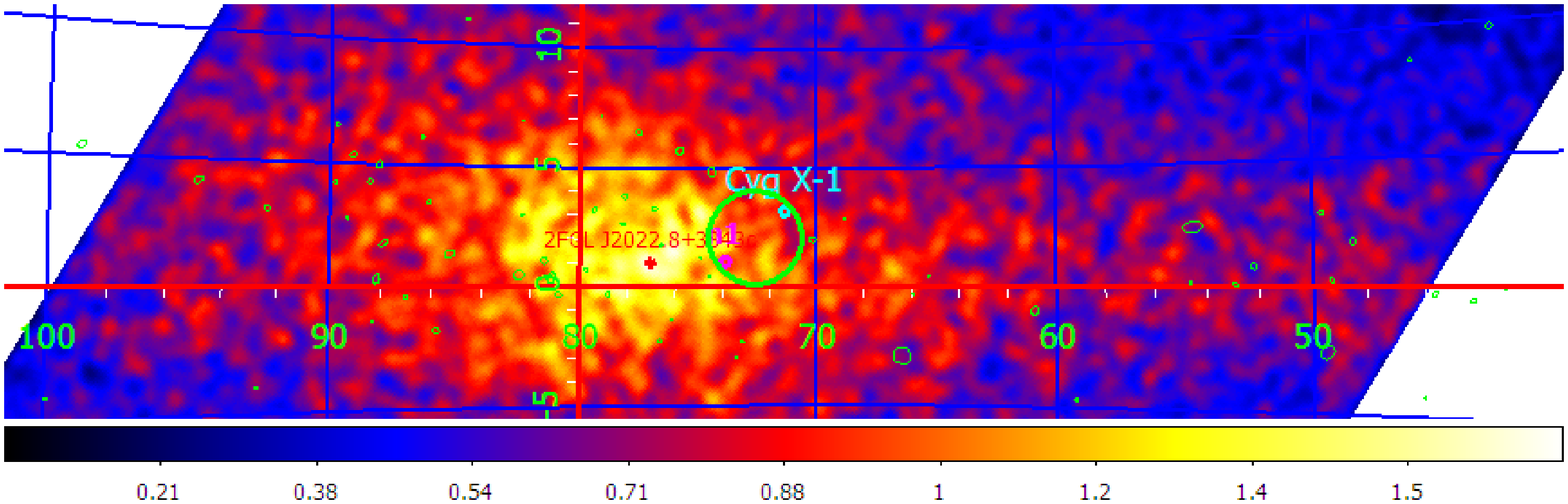}} 
\caption{The count-rate map for all the data in the 30--50 MeV energy range. The positions of Cyg X-1, {\tt n1}, and Cyg OB3 association are marked as in Fig.\ \ref{ts_maps}. The red point shows the position of the \fermi\/ source 2FGL J2022.8+3843c, corresponding to the SNR G076.9+01.0.
} \label{30_50}
\end{figure*}

We have analysed the data from the \fermi\/ LAT from the direction of Cyg X-1. We have performed binned \fermi/LAT data analysis using the v9r27p1 \fermi\/ Science Tools with {\tt P7SOURCE\_V6 IRF}. For the analysis, we have considered a $20\degr\times 20\degr$ region around Cyg X-1. We have included in the modelling of the region all sources from the 2-year \fermi\/ catalogue (2FGL) as well as the standard templates for Galactic (\texttt{gal\_2yearp7v6\_v0.fits}) and extragalactic (\texttt{iso\_p7v6source.txt}) backgrounds. The spectra of all catalogue sources were modelled with a power law. At the initial stage of the analysis, we have built the test-statistic (TS; \citealt{mattox96}) maps of the region at energies 0.3--1~GeV and 1--3~GeV, see Fig.\ \ref{ts_maps} (top). These maps show the significance ($\propto\sqrt{\rm TS}$) of a point-like source added to each point of the map. The 0.3--1 GeV map reveals a broad residual structure, which covers at the edge the position of Cyg X-1. In order to account for it, we have to introduce to the model an additional diffuse source with a constant flux within a $2\degr$ radius, shown by the green circle at Fig.\ \ref{ts_maps}. This source, centred on RA 301.51, Dec 35.72, approximately coincides with the Cyg OB3 stellar association (which centre is at RA 301.75, Dec 35.9). An additional point-like source marked as {\tt n1} (RA 303.42, Dec 36.21, ${\rm TS}\simeq 50$, 1--3 GeV) is shown with the small magenta circle. Its possible identification is the supernova remnant SNR G073.9+00.9 (RA 303.40, Dec 36.12; work in preparation). In our analysis, we use these sources in order to compensate for the residuals above the standard model of \fermi\/ Galactic diffuse background. Then, Cyg X-1 is detected with the model described above at the TS value of 14.8 (1--3~GeV), which corresponds to a $\simeq 4\sigma$ detection. 

We have then divided the data into the hard and soft state, based on light curves from the \xte\/ All-Sky Monitor (ASM; \citealt*{brs93,levine96}), the \swift\/ Burst Alert Telescope (BAT; \citealt{barthelmy05,m05}), and MAXI \citep{matsuoka09}. We have normalized the ASM and BAT count rates to the average values during the long hard state of MJD 53880--55375 \citep{z11b}. In the case of MAXI, we normalized its light curve to that of the ASM during their overlap. The light curves for the period studied here are shown in Fig.\ \ref{lc}. We have also considered the \fermi/GBM occultation data \citep{wh12}, but they closely follow the BAT data, though with larger error bars. Thus, we do not show them here.

By definition, the hard state in Fig.\ \ref{lc} corresponds to both the ASM and BAT fluxes around their respective averages. From that, we identify the hard-state MJD intervals as those of up to 55350, 55676--55790 and 55900--55940. The soft state corresponds to high ASM fluxes and low BAT fluxes, for which we find MJD 55390--55670, 55800--55890 and 55945--56020, 56100--56456 (the last day of the analysed data. This corresponds to the effective exposure for the hard and soft states of 822 d and 575 d, respectively. We do not detect Cyg X-1 in the soft state, and in the hard state, the detection significance is slightly higher (TS $=15.6$) than that in all of the data, see Fig.~\ref{ts_maps} (bottom). The low significance of the source as well as the presence of nearby residual structures with similar significance (see top right panel of Fig.~\ref{ts_maps}) makes a spurious nature of the detection possible. In our modelling, we treat the obtained flux values mostly as upper limits; however, we point out that the dependence of the flux on the source state (soft vs.\ hard) and its spatial coincidence with Cyg X-1 position make the reality of the detection rather likely.

For spectral analysis, we split the 0.1--300 GeV range into 7 logarithmically spaced bins. We fit the model described above in each bin separately. In this model, we fix the photon power law indices of all point sources and the added diffuse one to $2$, leaving the normalization free.

We have also analysed the data at the energies of 30--100 MeV, splitting it into the 30--50 MeV and 50--100 MeV bins. We consider the model as above except for replacing the standard isotropic background (given at energies $\leq 68$ MeV only) with its power law extrapolation to lower energies. Fig.\ \ref{30_50} shows the count-rate map for all of the data in the 30--50 MeV energy range. The brightest diffuse source in this energy range is clearly shifted from Cyg X-1, by $\approx 6\degr$, and its position corresponds to the SNR G076.9+01.0 (= \fermi\ source 2FGL J2022.8+3843c), which is shown by the red point.

\begin{figure}
\centerline{\includegraphics[width=\columnwidth]{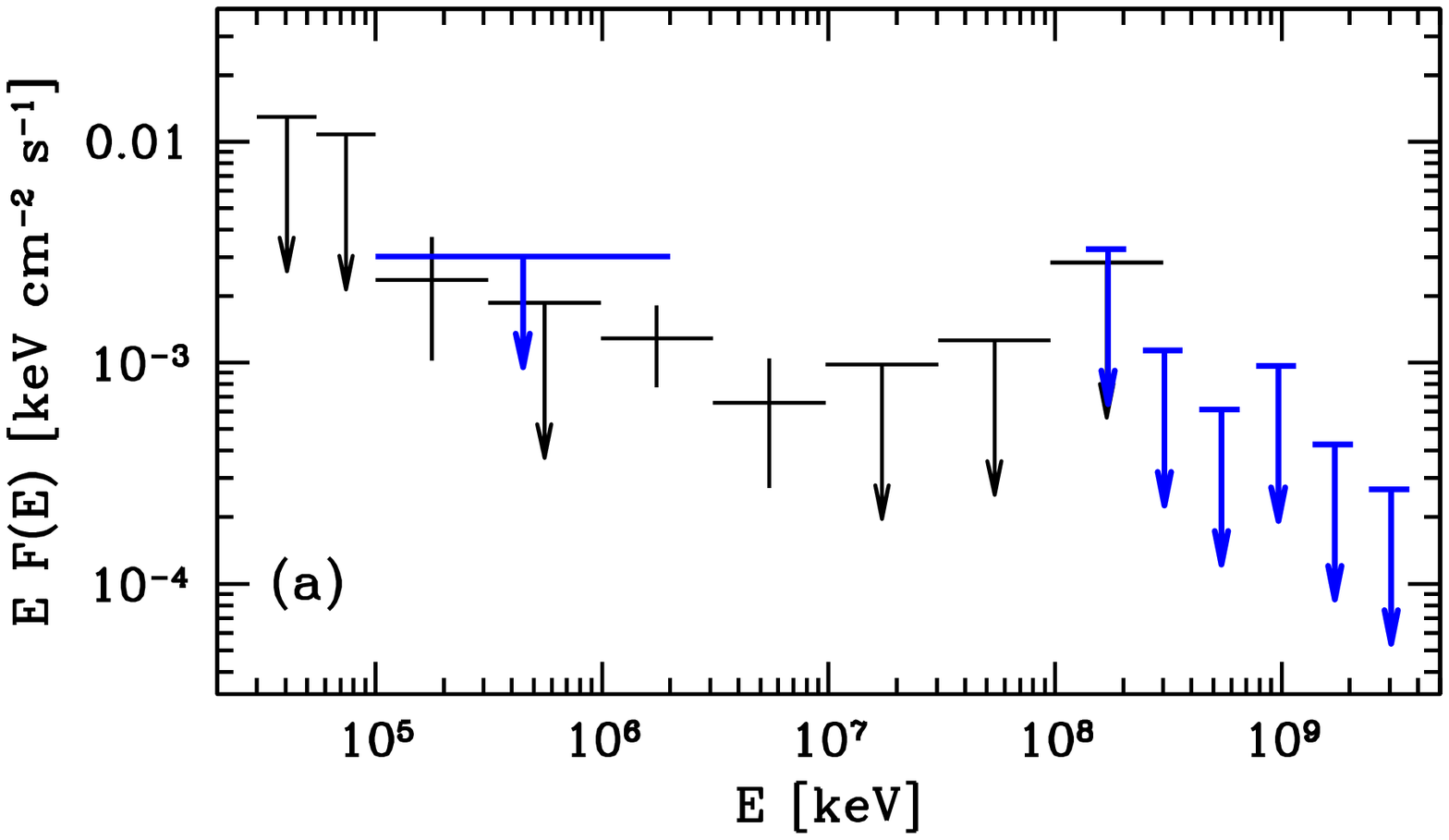}} 
\centerline{\includegraphics[width=\columnwidth]{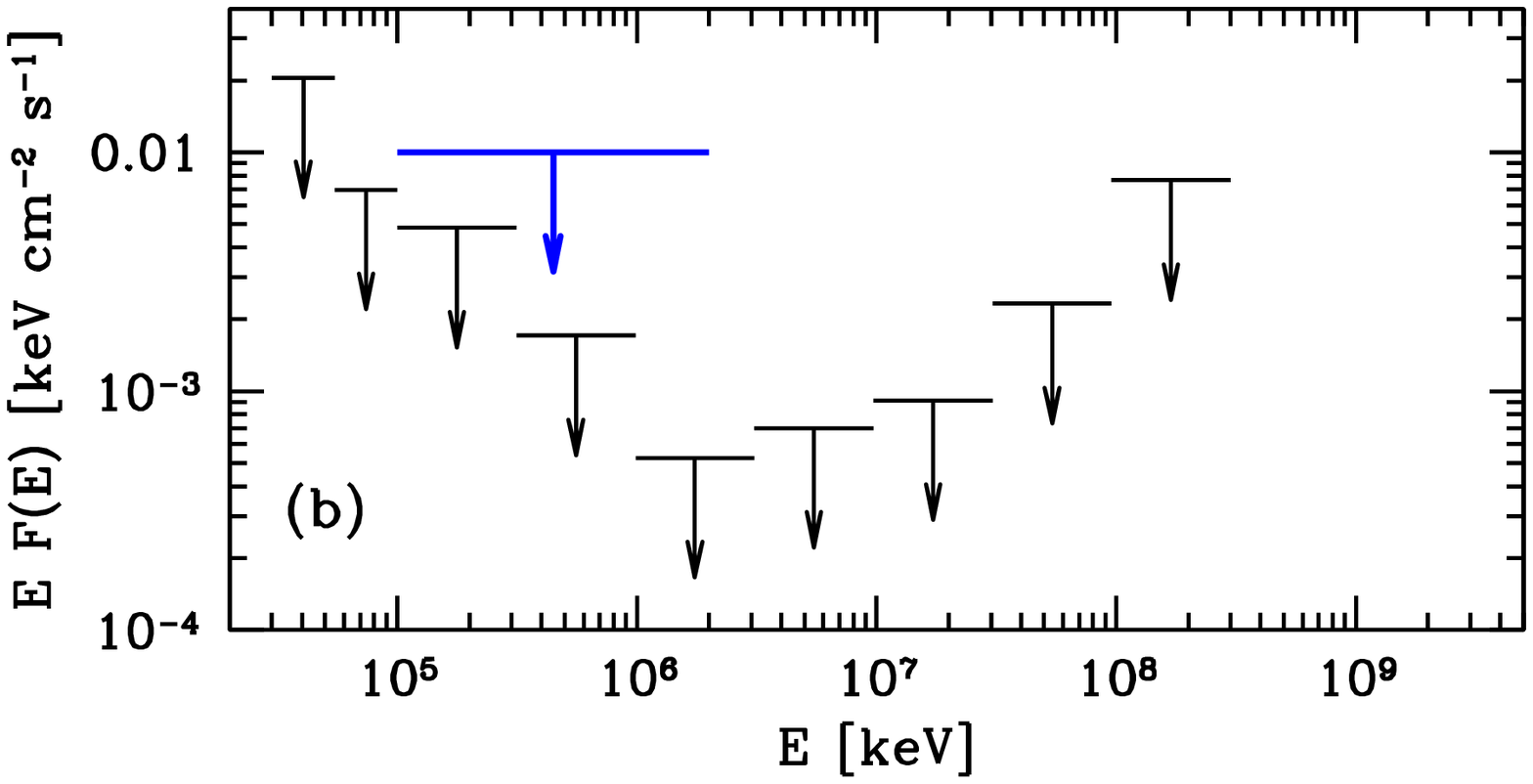}} 
\centerline{\includegraphics[width=\columnwidth]{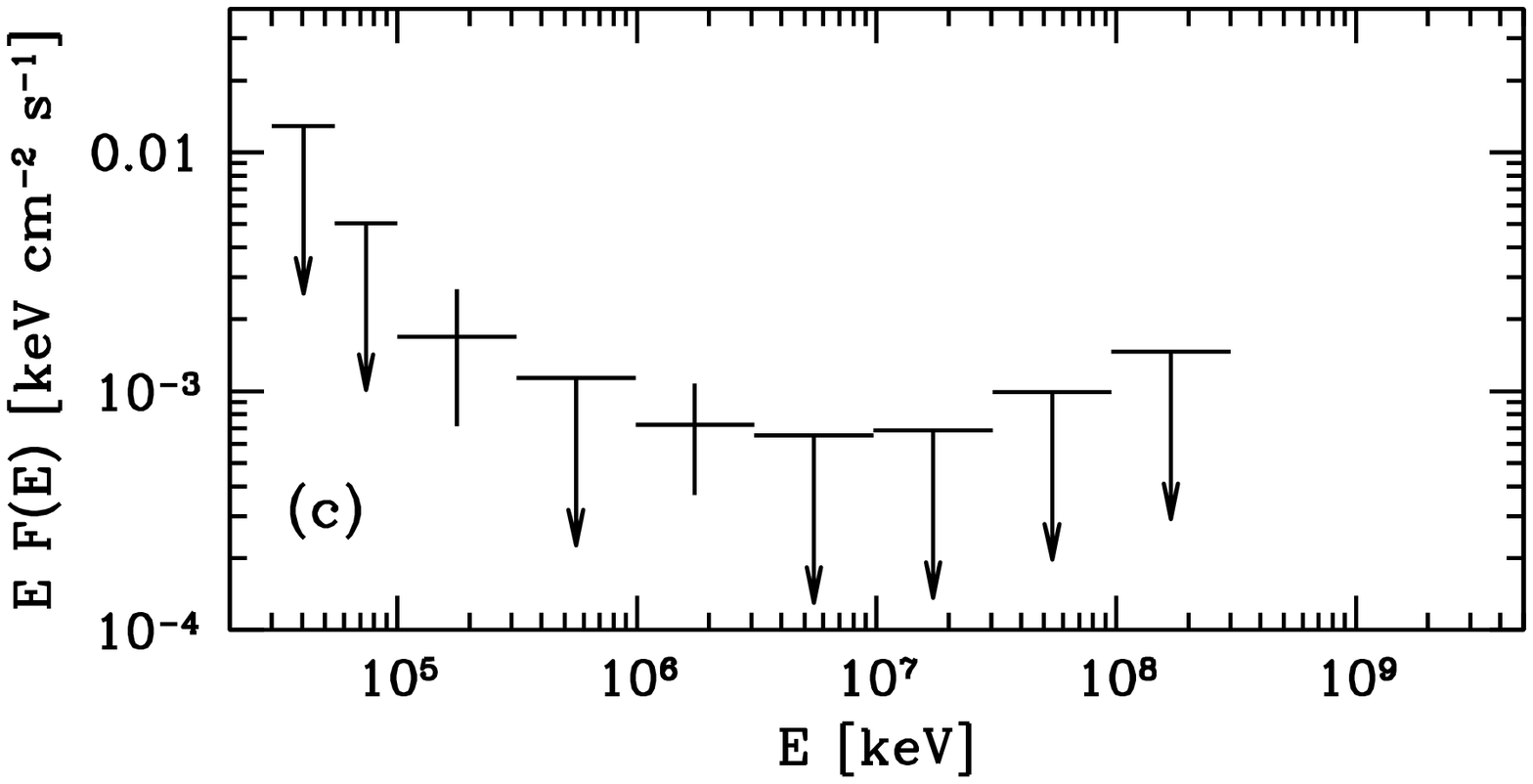}} 
\caption{The \fermi\/ LAT upper limits and measurements for Cyg X-1. The panels (a), (b) and (c) correspond to the hard and soft spectral states and to the entire analysed data, respectively. See Section \ref{fermi} for details. The heavy blue symbols show the upper limits in (a) from the MAGIC telescope \citep{magic} and in (a–-b), from \agile\/ \citep{sabatini10,sabatini13}. 
} \label{data}
\end{figure}

The obtained spectra and upper limits the hard and soft states and for all of the data are shown in Fig.\ \ref{data}. In the hard state, the 0.1--10 GeV spectrum can be fitted with a power law with the photon index of $\Gamma=2.57\pm 0.16$. Figs.\ \ref{data}(a) also show the upper limits obtained by the MAGIC Cherenkov telescope during the hard state \citep{magic}. We see they are at a similar $EF(E)$ level as our $\geq 3$ GeV upper limits. Fig.\ \ref{data}(a--b) also shows the previously obtained upper limits for the hard and soft states from \agile\/ of \citet{sabatini10,sabatini13} of $\simeq 3\times 10^{-3}$ keV cm$^{-2}$ s$^{-1}$, $\simeq 1\times 10^{-2}$ keV cm$^{-2}$ s$^{-1}$, respectively. Our upper limits and measurements are comparable in the hard state, but much lower in the soft state. 

We have also looked into a possible dependence of the hard-state emission on the orbital phase. Such a dependence is expected if a substantial part of the emission is due to Compton upscattering of the stellar blackbody photons \citep*{jackson72,dch10}. We have used the ephemeris of \citet{brocksopp99}. We have found the emission peaking during the first half of the orbital period, i.e., after the superior conjunction (black hole behind the donor; defining the 0 phase). This is consistent with the origin of the emission from Compton scattering of stellar photons in the jet (e.g., \citealt{dch10}), given that the position of the radio-emitting part of the jet lags behind that of the black hole, as evidenced by phase lags of the wind absorption of the radio emission \citep{sz07}. However, the detected \g-ray emission is weak and the statistical significance of the phase dependence is low. Thus, we do not present these results here. 

\section{Accretion models}
\label{accretion}

\subsection{Leptonic models}
\label{leptonic}

High-energy tails at $E\ga 1$ MeV in both hard and soft states of Cyg X-1 have been detected by the COMPTEL detector on board of \gro\/ (M02 and references therein) and by the \integral\/ SPI and IBIS detectors (e.g., \citealt{jrm12}; ZLS12). They have been often modelled by hybrid Comptonization (\citealt{av85,pc98,gierlinski99}; M02; \citealt{pv09}, hereafter PV09; \citealt{mb09,delsanto13}). In these models, the steady-state electron distribution in the presence of acceleration consists of a Maxwellian part and a high-energy, power-law like, tail. The high-energy tail above the thermal-Comptonization spectral component (peaking around 200 keV) in the hard state is due to emission of the power-law electrons. The tail in the soft state above the disc-blackbody spectral component (peaking around 1 keV) is due to Comptonization by both thermal and non-thermal electrons, see, e.g., \citet{gierlinski99}. The main parameters of these models relevant here are the compactness, defined as $\ell\equiv L\sigma_{\rm T}/R m_{\rm e} c^3$ (where $L$ is the luminosity of the source, $R$ is its characteristic size, $\sigma_{\rm T}$ is the Thomson cross section and $m_{\rm e}$ is the electron mass), the power-law index at which the electrons in the source are accelerated, $\Gamma_{\rm inj}$, and the maximum Lorentz factor of the accelerated electrons, $\gamma_{\rm max}$. The value of $\gamma_{\rm max}$ determines the maximum possible energy of \g-rays from Compton scattering, $E_{\rm max}\sim \min(m_{\rm e}c^2\gamma_{\rm max}, 3 k T\gamma_{\rm max}^2)$, where $T$ is the maximum temperature of the accretion disc. However, this maximum energy may be not observed because of absorption in pair-producing collisions of \g-rays with blackbody disc photons, $\gamma\gamma\rightarrow {\rm e}^+{\rm e}^-$. The optical depth to this process, $\tau_{\gamma\gamma}$, is proportional to the compactness parameter, $\ell$. Thus, its value affects the position of the high-energy cutoff, which may then be lower than the maximum possible scattered energy determined by $\gamma_{\rm max}$. Finally, $\Gamma_{\rm inj}$ determines the slope of the high-energy tail due to Compton scattering by non-thermal, power-law like electrons. 

\begin{figure*}
\centerline{\includegraphics[width=13cm]{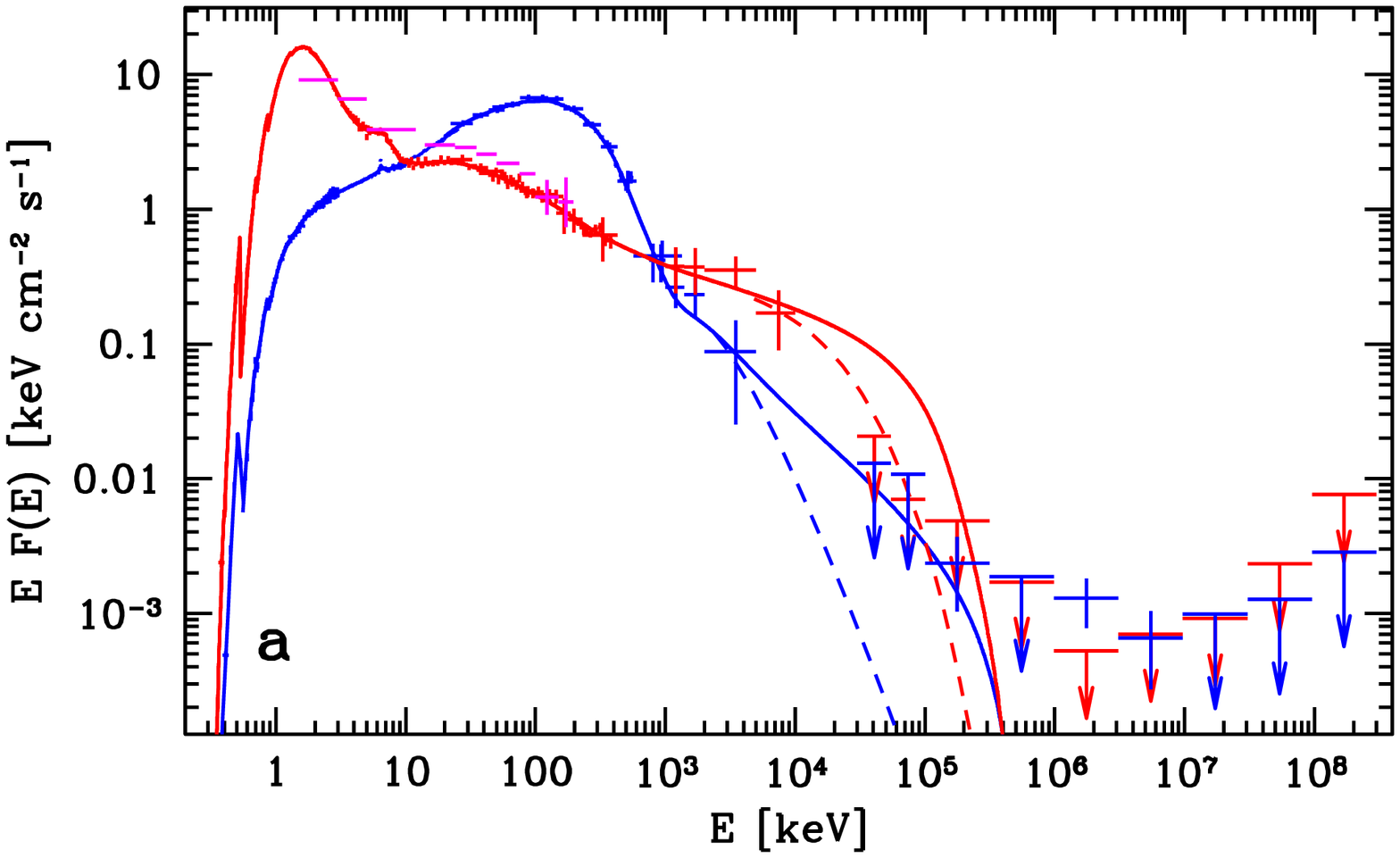}} 
\centerline{\includegraphics[width=13cm]{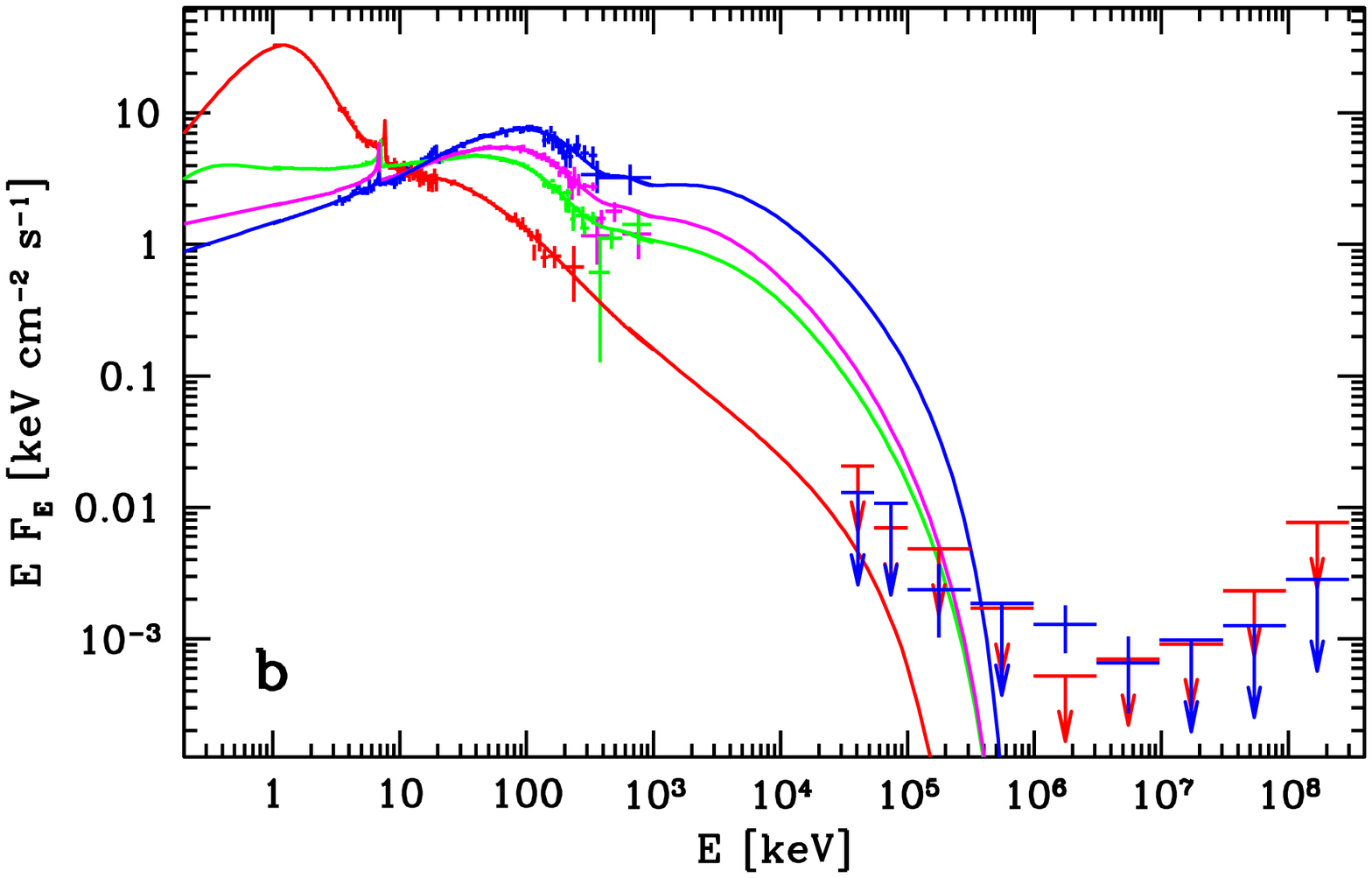}} 
\caption{(a) Broad-band X-ray/\g-ray data for Cyg X-1 in the hard (blue symbols) and soft (red and magenta symbols) states compared to hybrid-Comptonization accretion-flow models. The pointed-measurement data at $< 10$ MeV (attenuated by X-ray absorption) are from \sax\/ and \gro, and the data at $\geq 30$ MeV are from \fermi\/ (see Figs.\ \ref{data}a--b). The $< 10$ MeV data were fitted by hybrid Comptonization using the models of M02 and PV09 (solid and dashed curves, respectively). The magenta symbols show the average X-ray spectrum from the nearly simultaneous monitoring by the \xte\/ ASM and \swift\/ BAT. We see that this spectrum is in an overall agreement with the 1996 soft-state spectrum, in spite of corresponding to different time intervals. Due to the steepness of the $\ga 1$ MeV hard-state high-energy tail, the \fermi\/ upper limits  do not impose constraints on the accretion models in that state. However, the \fermi\/ soft-state upper limits in the 30--300 MeV range rule out the high-energy part of the model of M02. The difference between the two models causing the different high-energy behaviour is the assumed size of the source, $\sim 4\times 10^7$ cm ($\sim 20 R_{\rm g}$) in the model of PV09 and $\sim 4\times 10^8$ cm ($\sim 200 R_{\rm g}$) in the model of M02. (b) The \integral\/ data of \citet{delsanto13} fitted by their pure non-thermal (corrected for the X-ray absorption) model of \citet{mb09} together with the \fermi\/ data. We see that only the softest data set is compatible with that model. See Section \ref{accretion} for details.
} \label{hard_soft}
\end{figure*}

Fig.\ \ref{hard_soft}(a) shows the broad-band spectra in X-rays to soft \g-rays in the hard and soft states. The blue and red symbols correspond to the hard and soft state, respectively. The data at $\geq 30$ MeV are from \fermi, and are the same as those shown in Fig.\ \ref{data}(a--b). The data at $\leq 10$ MeV are the same as those shown in fig.\ 9 of M02, and are from the OSSE and COMPTEL detectors on board of \gro, supplemented at low energies by data from \sax\/ \citep{disalvo01,frontera01}. The OSSE and COMPTEL data in the hard state (blue crosses) represent the average spectrum in that state from \gro\/ observations. They have been fitted with the hybrid-Compton model {\tt eqpair} \citep{pc98,coppi99,gierlinski99} by M02, which spectrum is shown by the blue solid curve. In the hard state, the hybrid plasma probably forms an inner part of the accretion flow, overlapping with the optically thick disc \citep*{dgk07}. The blue dashed curve shows the analogous fit by PV09, who used a model which also incorporated synchrotron emission and absorption. All models shown here assume $\gamma_{\rm max}=10^3$ (arbitrarily chosen), which corresponds to $E_{\rm max}\simeq 0.5$ GeV, close to the maximum energy for the solid curves. The main difference between the hard state models is the fitted value of $\Gamma_{\rm inj}$, $=2.0$ for M02, and 3.8 for PV09, which results in a much steeper high-energy tail in the PV09 model compared to that of M02. Both models predict fluxes satisfying the constraints from \fermi. However, they are unable to explain the hard-state detection at 1--10 GeV.

The soft-state pointed measurements are from the 1996 soft state. Thus, they are from a single occurrence of that state. However, that form of the high-energy tail is known to vary from one occurrence of the soft state to another (e.g., \citealt{gz03,delsanto13}). Then, it is not clear whether the 1996 spectrum can be directly compared to the \fermi\/ soft-state upper limits, shown in Fig.\ \ref{hard_soft}. In order to test it, we show (magenta symbols) the average soft-state spectrum from the monitoring data by the ASM and BAT, simultaneous with each other (shown in fig.\ 10 of \citealt{z11b}). Those data overlap with the \fermi\/ observations. We see that the ASM/BAT spectrum is quite close to that of the 1996 soft state, with some of the differences attributable to different flux calibration. Thus, we use the fits to the 1996 soft state for comparison with the \fermi\/ upper limits. 

In the soft state, the hybrid plasma forms, most likely, localized coronal regions above an inner part of an optically-thick accretion disc, e.g., \citet{bkb76}, \citet{gierlinski99}, \citet{dgk07}. We see that the models of M02 and PV09 (red solid and dashed curves, respectively) differ in the high-energy cutoff, in spite of the same value of $\gamma_{\rm max}=10^3$ and the value of $\Gamma_{\rm inj}$ of PV09 being somewhat lower, 2.2, than that of M02, 2.6. This is an effect of the different values of the compactness, $\ell\simeq 3.7$ and 34 for M02 and PV09, respectively. The latter value results in a strong cutoff due to pair absorption, at an energy of $\sim (m_{\rm e}c^2)^2/(5 k T)\simeq 0.1$ GeV, which is seen the spectrum of PV09. The value of $\ell$ of PV09 corresponds to the size of the source of $R\simeq 4\times 10^7$ cm, which corresponds to $\sim 20 R_{\rm g}$, where $R_{\rm g}=GM_{\rm X}/c^2$ is the gravitational radius. This size approximately corresponds to the radius of the maximum of the gravitational energy release in an optically-thick accretion disc in the Schwarzschild metric. Thus, the lack of a detection of photons at $E>30$ MeV is consistent with the standard accretion scenario in the soft state. A similar result has been found by \citet{sabatini13} using the \agile\/ soft-state upper limit.

Fig.\ \ref{hard_soft}(b) shows some of the data and models of \citet{delsanto13} together with the \fermi\/ data. \citet{delsanto13} analyzed 12 \integral\/ data sets chosen based on the X-ray hardness. They fitted those data with the hybrid Comptonization models of both \citet{coppi99} and \citet{mb09}. Fig.\ \ref{hard_soft}(b) shows four of those sets, including the softest and the hardest, fitted with a model of \citet{mb09} with pure non-thermal acceleration/injection with $\gamma_{\rm max}=10^3$, $\ell=10$, and relatively hard $\Gamma_{\rm inj}\simeq 2.2$--3  (see the other parameters in table 4 of \citealt{delsanto13}; note that the normalization of the $\nu L_\nu$ spectra shown in their fig.\ 10 need to be increased by factors of $\simeq 12$, 5, 5, 6). We see that the softest spectrum is compatible with the pure non-thermal injection model. On the other hand, the remaining three data sets predict the 30--300 MeV fluxes significantly above the \fermi\/ data. Increasing the compactness would reduce these fluxes. However, the $\sim$1--3 MeV fluxes are much above the average of the hard state (\citealt{jrm12}; ZLS12), which indicates that models with pure non-thermal injection and hard $\Gamma_{\rm inj}$ are not compatible with the hard state data. \citet{delsanto13} have also fitted the hardest data with pure thermal models, which, given the fitted temperature of 70--80 keV, satisfy the \fermi\/ constraints.

\subsection{Hadronic models}
\label{hadronic}

Our upper limits also constrain the energies and densities of protons and He nuclei in the accretion flow. Radiatively inefficient hot accretion models, which may correspond to the hard state, do predict the presence of hot ions, which collisions lead to production of pions, whose decay, in turn, leads to substantial fluxes of \g-rays \citep*{mnk97,mahadevan99,om03,nxs13}. The models of \citet{mnk97}, \citet{mahadevan99}, and \citet{om03} predict \g-ray $EF(E)$ fluxes at levels similar or higher than those in X-rays. In Cyg X-1, this is strongly ruled out by the data. However, most of those models correspond to values of $\dot m\equiv \dot M c^2/L_{\rm E}$ (where $L_{\rm E}$ is the Eddington luminosity) much lower than that likely to correspond to Cyg X-1, and thus we cannot compare them directly to our data. 

The recent work of \citet{nxs13} considers the case of $\dot m=0.1$, which is much closer to the case of Cyg X-1, which has the average $L/L_{\rm E}\simeq 10^{-2}$ in the hard state (e.g., \citealt{z02}), which corresponds to $\dot m\simeq 0.1$ for its likely accretion efficiency of $\sim 0.1$ \citep{mbf09}. In the hard state of Cyg X-1, $EF(E)$ around 2--10 keV is $\simeq 2$ keV cm$^{-2}$ s$^{-1}$, whereas that at 0.1--10 GeV is $\la 2 \times 10^{-3}$ keV cm$^{-2}$ s$^{-1}$. The resulting ratio of $\simeq 10^{-3}$ is similar to that found by \citet{nxs13} for the model with thermal protons and 1/2 of the electrons viscously heated (which large fraction is required in Cyg X-1, given its large accretion efficiency). Thus, our data constrain the proton distribution in the accretion flow to thermal or quasi-thermal.

On the other hand, the efficiency of the models of \citet{nxs13} is $\simeq 0.02$, 0.1 for the black-hole spin of $a_{\rm BH}=0$ and 0.998, respectively. The efficiency for $a_{\rm BH}=0$ is lower than the presumed $\sim 0.1$ for Cyg X-1. Thus, a model with $\dot m$ larger than 0.1 would apply, which would have photon-photon pair absorption stronger than that found in \citet{nxs13}. This would, in turn, reduce the ƒ\g-ray luminosity relative to the X-ray one and would allow for models with some fraction of the protons being non-thermal.

\section{Jet models}
\label{jet}

\begin{figure*}
\centerline{\includegraphics[width=14cm]{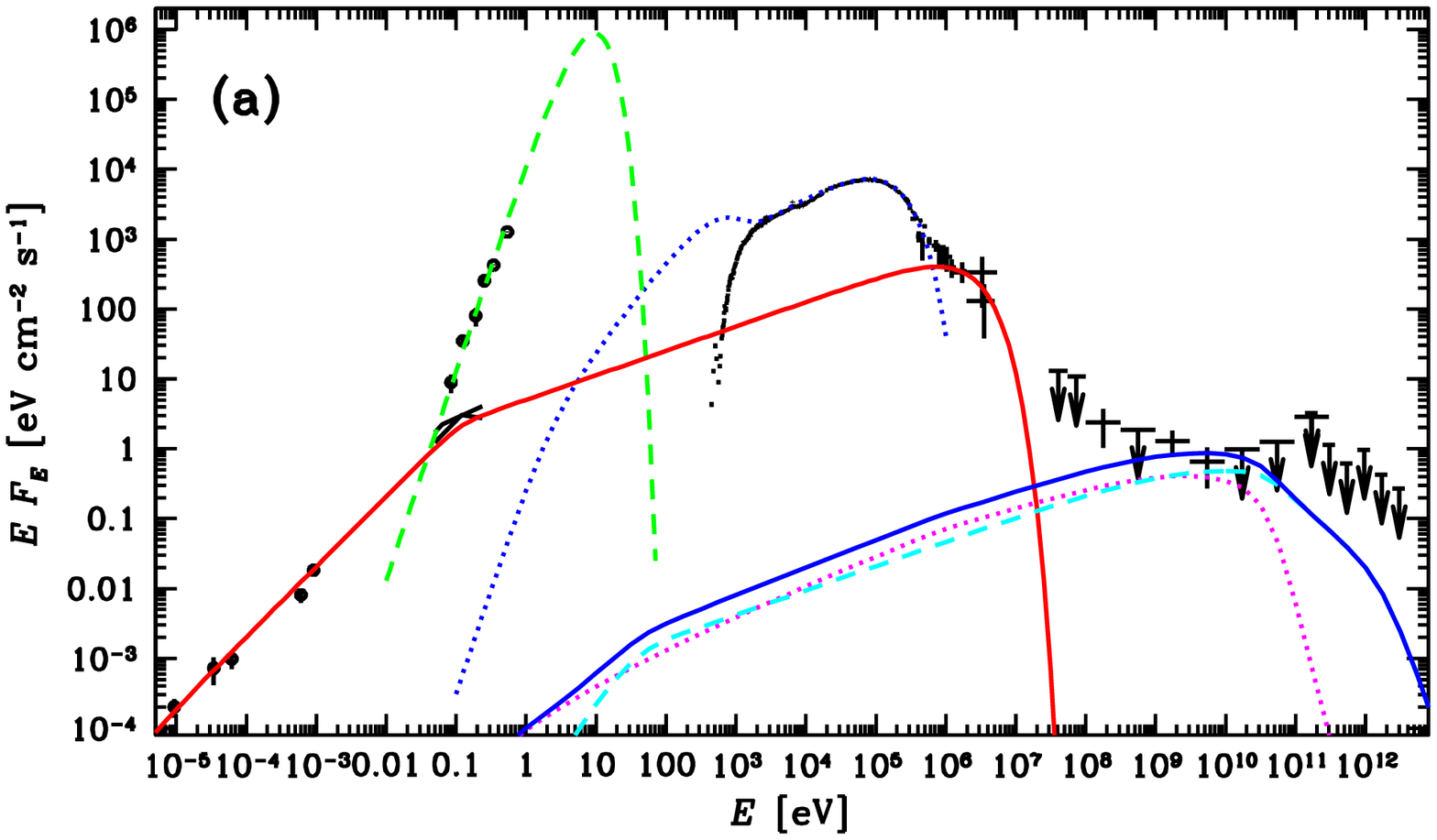}} 
\centerline{\includegraphics[width=14cm]{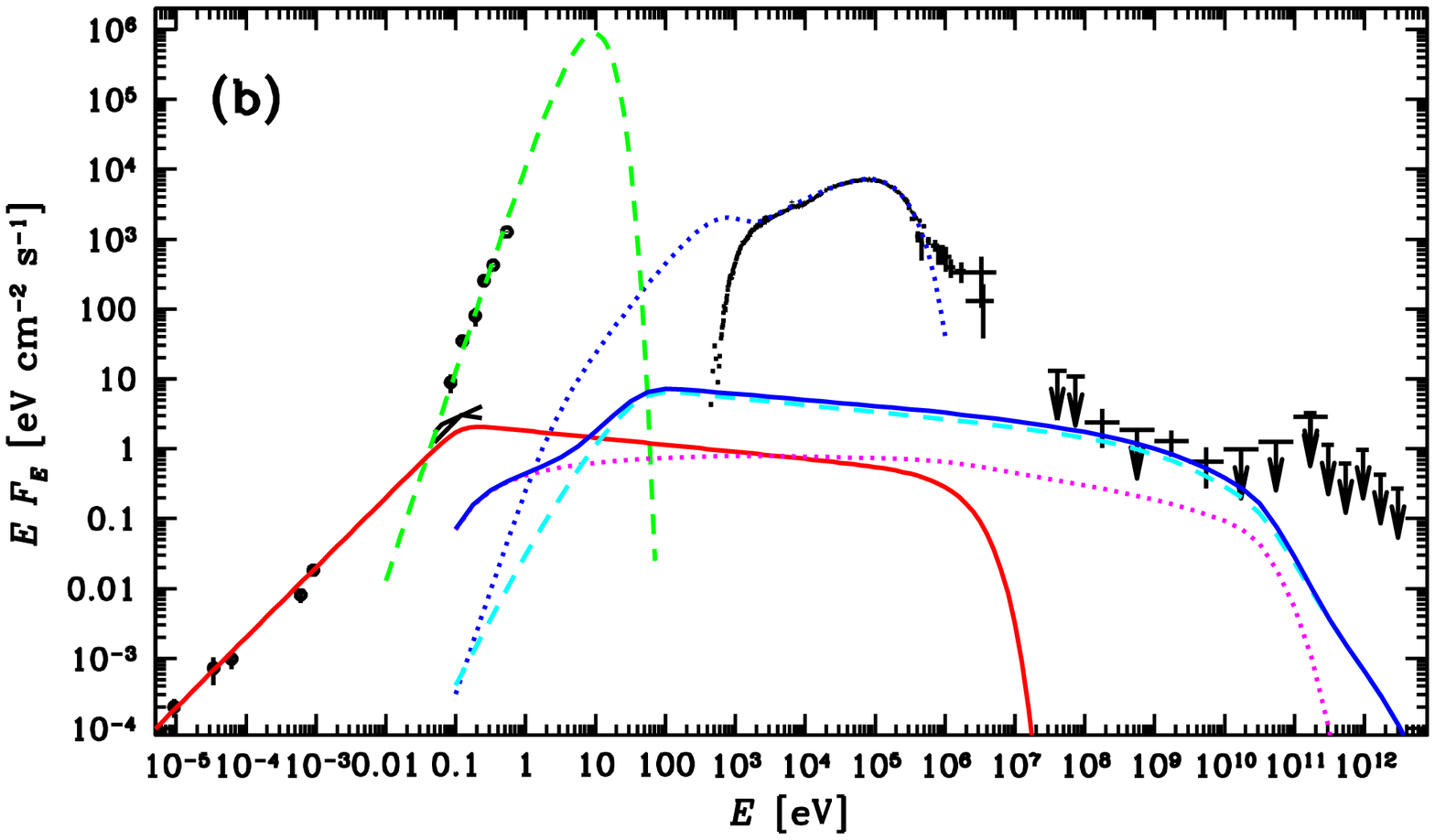}} 
\caption{The average hard-state radio to \g-ray spectrum (black symbols) of Cyg X-1 and its donor shown together with model spectra. The data up to 5 MeV are the same as those in ZLS12, the 30 MeV--300 GeV symbols give the results of this work, see Fig.\ \ref{data}(b), and the the 6 upper limits at the highest energies are from MAGIC. The green dashed curve show the stellar blackbody. The dotted blue curve shows an estimated unabsorbed accretion disc and hot flow model, see \citet{zps13}. The shown spectral components are from the jet model described in Section \ref{jet}, which assumes a single power-law electron distribution. The electron index is (a) $s=2.3$, accounting for the observed MeV tail (claimed to be strongly polarized), which corresponds to the approximately maximum jet emission allowed by the data, and (b) $s=3.2$. The red solid, magenta dotted and cyan dashed curves show the model synchrotron, synchrotron self-Compton and blackbody-Compton components, respectively. The solid blue curves give the sum of the two Compton spectra. 
} \label{jet_model}
\end{figure*}

A strong steady radio jet \citep{stirling01} is present in Cyg X-1 in the hard state, though there is also a weaker and variable radio emission in the soft state \citep{z11b,rushton12}. Here we consider the implications of the presence of the jet in the hard state only. We consider only leptonic models, in which the GeV emission is due to Compton upscattering by the jet relativistic electrons. A more detailed study of the leptonic jet models in the context of the \fermi\/ measurements and upper limits is given in \citet*{zps13}. We also note that hadronic jet models for high-energy \g-ray emission of jets in high-mass X-ray binaries have been proposed by, e.g., \citet{romero03}, \citet{orellana07}, \citet*{boschramon05} and \citet{aharonian06}. In those models, the \g-ray emission is due to collisions of protons and He nuclei in the jet with either those of the stellar wind or interstellar medium or with accretion disc photons. In the case of interaction with ambient matter, the jet bulk velocity is sufficient for production of pions. In the case of interactions with photons, highly energetic protons, $\ga 100$ TeV, are required for photo-meson interactions. A study of those models is outside of the scope of this work. 

In studies of synchrotron jet emission, it is common to assume the electron steady-state distribution through the jet to be a single power law, e.g., \citet{bk79}, \citet{fb95} or \citet*{brp06}. This simple case, neglecting effects of radiative cooling on the shape of the distribution, may possibly correspond to the case of an acceleration process acting though the full jet volume. We thus assume a conical jet with the electron distribution, $N(\gamma)$, and magnetic field strength, $B$, given by, 
\begin{equation}
N(\gamma) = {K_0\gamma^{-s}\over \xi^2}{\rm e}^{-\left(\gamma/ \gamma_{\rm max}\right)^2},\quad \gamma_{\rm max}\simeq \left(9 B_{\rm cr}\xi \over 8\upi\alpha_{\rm f}\eta_{\rm acc} B_0\right)^{1/2},\quad B={B_0\over \xi},
\label{ngamma}
\end{equation}
where $\xi\equiv z/z_0$, $\gamma\geq\gamma_{\rm min}$, $s$ is the steady-state power-law index, $z$ is the height along the jet from the black-hole centre, $z_0$ corresponds to the onset of the emission, $\gamma_{\rm max}$ is the high energy cutoff following from the balance of synchrotron losses and acceleration on the time scale of $\eta_{\rm acc}$ of the Larmor periods, $\alpha_{\rm f}$ is the fine-structure constant, $B_{\rm cr}={2\upi m_{\rm e}^2 c^3/e h}$ is the critical magnetic field, $e$ is the electron charge, $h$ is the Planck constant, and $K_0$ is the normalization. The adopted $N(\gamma)$ has a constant electron number per unit $\gamma$ and unit length. The energy index of the optically-thin power-law part of the spectrum is $\alpha= (s-1)/2$. 

We basically follow the method of ZLS12, in which jet partially self-absorbed synchrotron emission and synchrotron self-Compton (SSC) are taken into account. However, we also include Compton scattering of stellar blackbody photons (hereafter BBC), for which we take into account the angle-dependent Klein-Nishina cross section \citep{aa81}, following the formalism of \citet{zps13}. The normalization, $K_0$, is determined in our modelling by the average flux at 15 GHz, which we adopt as 13 mJy. Our formalism also includes pair absorption of \g-rays on stellar photons, taking into account the finite size of the star \citep{bednarek97}. Both the BBC flux and  $\tau_{\gamma\gamma}$ strongly depend on the orbital phase, with both having maxima at the superior conjunction. Here, we calculate the BBC flux and $\tau_{\gamma\gamma}$ at each phase and compute the average absorbed spectra.

Fig.\ \ref{jet_model}(a) shows a model in which the MeV tail is attributed to synchrotron jet emission, as claimed by \citet{jourdain12}. The index of the steady-state electron distribution is $s=2.3$. If this index is due to acceleration and subsequent cooling, the electrons are to be accelerated at a rather hard power law, with the index of $\Gamma_{\rm inj}=1.3$. This is required to account for both the flux at the infrared turnover frequency (claimed by \citealt{rahoui11}) and the MeV flux, see ZLS12 for details. We note that if the emitting electrons are efficiently cooled in a part of the jet, the synchrotron cooling rate decreasing with height due to the decrease of $B$ leads to $N(\gamma)\propto \xi^{-1}$ at some values of $\xi$ and $\gamma$, different from the dependence of equation (\ref{ngamma}). This possibility is not taken into account here. To account for the steep slope of the MeV tail, $\eta_{\rm acc}\simeq 20$ is required, which value we adopt. 

The optically-thin synchrotron flux is $\propto K_0 B_0^{(s+1)/2}$. Thus, a given flux can be obtained for either a low $K_0$ and high $B_0$ or vice versa. On the other hand, the rate of Compton scattering is $\propto K_0$. Thus, $K_0$ has to be sufficiently low for the Compton-scattered components to yield fluxes $\leq$ the \fermi\/ data points. This results in a lower limit on $B_0$. Then, equation (22) of ZLS12 relates $B_0$ to $z_0$. In the present model, we modify this formula slightly by taking into account the finite extent of the jet, for which we assume $z_{\rm max}=10^{15}$ cm \citep{stirling01}. In the model shown in Fig.\ \ref{jet_model}(a), we obtain $B_0=4\times 10^4$ G and $z_0 \simeq 2.6\times 10^9$ cm ($\simeq 1.1\times 10^3 GM/c^2$). These parameters imply that cooling is important for all electrons emitting synchrotron radiation above the turnover energy. We note that the resulting model yields the 0.1--0.3 GeV flux a factor of several below the LAT measurement.

The magnetic field is rather high in order to satisfy the GeV measurements, and it is found to be strongly above equipartition with the electron pressure, with the plasma parameter of $\beta=(u_{\rm e}/3)/(B^2/8\upi)\simeq 0.008$, where $u_{\rm e}$ is the pressure of the relativistic electrons. On the other hand, the magnetization parameter is $\sigma\simeq (B^2/8\upi)/w$, where $w$ is the enthalpy. If the ions are cold protons, $w\simeq n_{\rm p}m_{\rm p}c^2+(4/3)u_{\rm e}$, where $n_{\rm p}$ and $m_{\rm p}$ is the proton number density and mass, respectively. We then define $\eta_{\rm p}=n_{\rm p}/n_{\rm e,rel}$, where $n_{\rm e,rel}$ is the number density of the accelerated relativistic electrons. If not all electrons are accelerated and in the absence of pairs, $\eta_{\rm p}> 1$. In the present model, we find $\sigma$ to be small, $\sigma\simeq 0.1/\eta_{\rm p}$. Thus, the jet is not magnetically dominated. The jet+counterjet power [see equations (33--35) in ZLS12] in the relativistic electrons, the protons, the magnetic field and the synchrotron power emitted in all directions in the optically-thin part of the spectrum is $P_{\rm e}\simeq 3.1\times 10^{33}$ erg s$^{-1}$, $P_{\rm p}\simeq 2.1\,\eta_{\rm p}\times 10^{35}$ erg s$^{-1}$, $P_{\rm B}\simeq 9.5\times 10^{34}$ erg s$^{-1}$, and $P_{\rm S}\simeq 4.5\times 10^{35}$ erg s$^{-1}$, respectively. $P_{\rm e}\ll P_{\rm S}$ implies that the electrons have to be efficiently reaccelerated. Also, $\eta_{\rm p}>1$ is required for $P_{\rm S}<P_{\rm e}+P_{\rm p}+P_{\rm B}$. 

We then consider a model with $s=3.2$, which may correspond to the acceleration index of $\Gamma_{\rm inj}\simeq 2.2$, which is a typical value for acceleration processes. That model, shown in Fig.\ \ref{jet_model}(b), does not account for the MeV tail, which then is presumed to be due to hybrid Comptonization in the accretion flow (Section \ref{leptonic}). Its parameters are $B_0=2.5\times 10^3$ G and $z_0\simeq 2.0\times 10^9$ cm ($\simeq 8.3\times 10^2 GM/c^2$). This model reproduces well the LAT data. The model has the Lorentz factor corresponding to emission at the turnover frequency of $\gamma_{\rm t}\simeq 55$, and the cooling break Lorentz factor in the jet region dominated by synchrotron losses of $\gamma_{\rm b}\simeq 10^3\xi$. The GeV emission is produced by electrons with $\gamma\sim 10^4$--$10^5$. Thus, there would be a cooling break at intermediate energies in a part of the jet, which effect is not included in our simple model. 

Given the steep electron distribution, the degree of equipartition and the jet powers depend strongly on $\gamma_{\rm min}$. For self-consistency, $\gamma_{\rm min}<\gamma_{\rm t}$ is required. For $\gamma_{\rm min}= 50$, $\beta\simeq 65$, $\sigma\simeq 2.4\times 10^{-4}/\eta_{\rm p}$. Thus, the jet is much below equipartition and strongly matter-dominated. For a stronger magnetic field, the GeV emission would be below the \fermi\/ data. An equipartition model with an index between the two considered cases, $s=2.3$ and $3.2$, would yield the observed GeV flux. At $\gamma_{\rm min}=50$, $P_{\rm e}\simeq 5.3\times 10^{34}$ erg s$^{-1}$, $P_{\rm p}\simeq 1.6\,\eta_{\rm p}\times 10^{35}$ erg s$^{-1}$, and $P_{\rm B}\simeq 2.1\times 10^{32}$ erg s$^{-1}$. In this model, $P_{\rm S}\simeq 6.5\times 10^{33}$ erg s$^{-1}$ and the power in the BBC component is $P_{\rm C}\simeq 2.1\times 10^{34}$ erg s$^{-1}$. Then, the radiative output is $\sim$0.1 of the jet power. If $\gamma_{\rm min}= 1$, the kinetic power in the electrons and protons become even larger, $P_{\rm e}\simeq 5.8\times 10^{36}$ erg s$^{-1}$, $P_{\rm p}\simeq 8.7\,\eta_{\rm p}\times 10^{38}$ erg s$^{-1}$. The kinetic power would then be much larger than the accretion radiative power. This argues for $\gamma_{\rm min}\gg 1$. 

In both models, the BBC process is important, and it dominates over the SSC component in the second model. The bulk of the BBC emission (i.e., the maximum of ${\rm d}F/{\rm d}\,\ln z$) is found to be from $z\simeq 10^3 z_0$, which is of the order of the binary separation, $a$. At this height, the optical depth to pair absorption in collisions of BBC \g-rays with stellar blackbody photons becomes $>1$ for $E\ga 10^{11}$ eV. Since the jet emission at such energies is both weak and below the upper limits, pair absorption is only marginally important for our models. The bulk of the SSC emission originates from heights much below the binary separation, which implies that this emission is absorbed somewhat more than the BBC one.

\section{Discussion}
\label{discussion}

We have detected high-energy \g-rays from Cyg X-1 in the hard state, but not in the soft state. This, on the surface, might appear contrary to the case of another high-mass X-ray binary, Cyg X-3, where the GeV emission has been detected only when the X-ray spectrum is soft \citep{fermi}. However, that state appears to be of a very-high type associated with the presence of a strong and flaring radio emission \citep*{szm08}. Cyg X-1 never enters that state, and its radio emission during the soft state is weak \citep{z11b,rushton12}. Thus, emission of high-energy \g-rays appears to be associated with the presence of significant radio emission. We note that Cyg X-3 in the hard state shows a relatively steady radio emission correlated with the X-ray flux (e.g., \citealt{szm08}), during which a steady \g-ray flux may be emitted, though it is difficult to separate it from a background diffuse emission \citep{fermi}. 

The issue of the high-energy \g-ray emission from Cyg X-1 in the hard state is strongly related to the origin of its MeV tail. There are currently two competing scenarios for its origin. One is hybrid Comptonization, the other is jet synchrotron emission. Our results confirm and reinforce those of ZLS12 (who neglected the BBC process, found to be important by us) that the jet accounting for the MeV tail has to have magnetic field strongly above equipartition. This implies that the accelerated electrons are in the fast cooling regime, which, in turn, requires a very hard acceleration index, $\Gamma_{\rm inj}\simeq 1.3$, which is not usual for acceleration processes. 

We note that a way to discriminate between the two scenarios may be to study short correlated time-scale variability between the thermal-Compton component, dominant below a few hundred keV, and the MeV tail. If both originate in an inner accretion flow, their variability should be strongly correlated down to ms time scale, though probably not directly proportional due to the likely presence of spectral variability. On the other hand, the jet emission around its base would respond to changes in the inner accretion flow on time scales $\ga z_0/c\sim 0.1$ s. Also, given that the jet emission is due to very different radiative processes than that in the accretion flow, the correlation may be much less strict. Furthermore, the component due to Compton scattering of stellar emission will react to changed condition in the inner accretion flow on time scales of $\ga a/c\simeq 100$ s.

In the soft state, we have found that our \fermi\/ upper limits imply that the size of the region emitting the high-energy tail has to be at most a few tens of $R_{\rm g}$. Then, strong pair absorption of \g-rays in collisions with disc blackbody photons yields a cutoff compatible with our upper limits. We note, however, that the corona can be larger if the maximum Lorentz factor of the electrons is low, $\la 500$, in which case there will be no Compton emission at energies $\ga 0.1$ GeV.

\section{Conclusions}
\label{conclusions}

We have obtained measurements and upper limits of the flux from Cyg X-1 in the 0.03--300~GeV energy band based on observations of the \fermi/LAT. We have detected a steady emission at a $4\sigma$ significance in the 0.1--10~GeV energy band in the hard spectral state. That emission can be approximately described as a power law with $\Gamma\simeq 2.6\pm 0.2$. On the other hand, we have found only upper limits in the soft spectral state. Our measurements and limits are significantly below previous upper limits from \agile. 

We have studied implications of our measurements for accretion and jet models. In the soft state, the upper limits imply the size of hot corona in the accreting source to be $\la 20 R_{\rm g}$, which results in strong attenuation of the non-thermal emission at $\ga 0.1$ GeV by pair absorption in collisions of \g-rays with blackbody disc photons. In the hard state, our measurements rule out most of the published hadronic accretion models, in which the GeV emission is due to decay of neutral pions produced in ion-ion collisions. Some hadronic models, however, appear compatible with the data. 

In the hard state, the recent claims of very strong linear polarization in the MeV range imply the jet synchrotron emission dominates that energy band. We find we can fit the data with a jet model. Compared to the previous model of ZLS12, we have taken into account Compton upscattering of stellar photons, which has been found to dominate the GeV emission. This reinforces the conclusion of ZLS12 that such models require the jet magnetic field to be strongly above equipartition. The strong magnetic field implies the fast-cooling regime, which in turn implies that electrons in the jet have to be accelerated at a hard power-law index. On the other hand, we find we can explain the hard-state GeV emission by a jet model with a softer acceleration rate, in which case the MeV tail is explained by hybrid Comptonization in the accretion flow, and not by the jet.

\section*{ACKNOWLEDGMENTS}

We thank M. Sikora, A. Nied{\'z}wiecki, J. Malzac and A. Strong for discussions and comments, P. Pjanka for assistance with the jet numerical model,  J. Miko{\l}ajewska for pointing out the possible identification of {\tt n1} with SNR G073.9+00.9,  and B. Cerutti, the referee, for valuable comments. We thank M. Del Santo and J. Poutanen for providing us with the data used in \citet{delsanto13} and PV09, respectively. DM and MC would like to thank the participants of the ISSI team ``Study of Gamma-ray Loud Binary Systems" for useful discussions. The work of DM has been supported in part by the Ukrainian NAS program of space research 2013 and by the State Program of Implementation of Grid Technology in the Ukraine. AAZ has been supported in part by the Polish NCN grants N N203 581240 and 2012/04/M/ST9/00780. The authors also wish to acknowledge the SFI/HEA Irish Centre for High-End Computing (ICHEC) for providing computational facilities and support.

\label{lastpage}

\end{document}